\documentclass{emulateapj}
\usepackage{graphicx,url,amssymb,amsmath,longtable,rotating,color,units,wasysym,subfigure,epsfig,listings,bm}

\slugcomment{Submitted for publication in The Astrophysical Journal}
\shorttitle{GWs from Fallback Accretion onto NSs}
\shortauthors{Piro and Thrane}

\newcommand{\be}{\begin{eqnarray}}
\newcommand{\ee}{\end{eqnarray}}
\newcommand{\lp}{\left(}
\newcommand{\rp}{\right)}
\newcommand{\lb}{\left[}
\newcommand{\rb}{\right]}

\begin{document}


\title{Gravitational Waves from Fallback Accretion onto Neutron Stars}

\author{Anthony L. Piro\altaffilmark{1} and Eric Thrane\altaffilmark{2}}

\altaffiltext{1}{Theoretical Astrophysics, California Institute of Technology, 1200 E California Blvd., M/C 350-17, Pasadena, CA 91125; piro@caltech.edu}

\altaffiltext{2}{School of Physics and Astronomy, University of Minnesota, Minneapolis, MN 55455, USA; eric.thrane@ligo.org}


\begin{abstract}
Massive stars generally end their lives as neutron stars (NSs) or black holes (BHs), with NS formation typically occurring at the low mass end and collapse to a BH more likely at the high mass end. In an intermediate regime, with a mass range that depends on the uncertain details of rotation and mass loss during the star's life, a NS is initially formed which then experiences fallback accretion and collapse to a BH. The electromagnetic consequence of such an event is not clear. Depending on the progenitor's structure, possibilities range from a long gamma-ray burst to a Type II supernova (that may or may not be jet-powered) to a collapse with a weak electromagnetic signature. Gravitational waves (GWs) provide the exciting opportunity to peer through the envelope of a dying massive star and directly probe what is occurring inside. We explore whether fallback onto young NSs can be detected by ground-based interferometers. When the incoming material has sufficient angular momentum to form a disk, the accretion spins up the NS sufficiently to produce non-axisymmetric instabilities and gravitational radiation at frequencies of $\unit[\sim700-2400]{Hz}$ for $\unit[\sim30-3000]{s}$ until collapse to a BH occurs. Using a realistic excess cross-power search algorithm, we show that such events are detectable by Advanced LIGO out to $\approx \unit[17]{Mpc}$. From the rate of nearby core-collapse supernovae, we estimate that there will be $\sim1-2$ events each year that are worth checking for fallback GWs. The observation of these unique GW signatures coincident with electromagnetic detections would identify the transient events that are associated with this channel of BH formation, while providing information about the protoneutron star progenitor.
\end{abstract}

\keywords{black hole physics ---
	gravitational waves ---
	stars: neutron ---
	supernova: general}


\section{Introduction}

Determining the fate of zero-age main-sequence (ZAMS) stars with large masses is a long-standing problem. In general, it is expected to depend in a complicated way on mass loss and rotation during the star's life. These in turn are related to details such as the magnetic field, metallicity, and binarity. Even with all these uncertainties, theoretical efforts indicate a rough general picture. A core-collapse supernova that successfully unbinds its stellar mantle leaves a neutron star (NS) behind. In cases when this does not happen, a stellar-mass black hole (BH) is instead expected, but this can occur in a number of different ways; \citep[for a more detailed discussion, see][]{oo11}. For example, if there is a nuclear phase transition during protoneutron star (PNS) cooling, or if cooling reduces pressure support in a hyper-massive PNS, a BH results. In another scenario, if the supernova mechanism fails to revive the accretion shock, continued accretion pushes the PNS over its maximum mass, creating a BH with likely little or no electromagnetic signal \citep{koc08}. Finally, if  the core-collapse supernova is successful, but perhaps weak, then the young NS will be subject to fallback accretion rates of $\dot{M}\sim10^{-4}-10^{-2}M_\odot\ {\rm s^{-1}}$ over the next $\sim30-1000\ {\rm s}$ \citep{mac01,zha08}. This additional material pushes the NS past its maximum mass, again resulting in a BH.

In the present work, we focus on this latter fallback mechanism for creating BHs. Since the idea of fallback accretion was first discussed by \citet{col71}, it has been an important area of focus for theoretical studies \citep[e.g.,][]{che89,ww95,fry99,zha08,ugl12}. There have also been a wide range of predictions for the type of events associated with fallback accretion leading to BHs. When rotation is included, the newly formed BH continues to accrete and may produce a jet. Depending on the mass of the envelope at the end of the star's life, this may result in a jet-powered Type II supernovae or a long gamma-ray burst \citep{mac01}. In cases where the jet is pointed away from the observer or jet formation does not occur, a dim supernova could result instead \citep{fry07,fbb09,mor10}. If a quickly-spinning, strongly-magnetized NS is present, it may fling away the infalling material and produce a ``propeller nova'' instead \citep{po11}.

Observationally, it would be helpful to determine the main-sequence mass range that leads to BHs created via fallback, so that it can be compared with theoretical expectations. \citet{heg03} argue that for sub-solar metallicities, this occurs for a ZAMS mass of $\sim25-40M_\odot$, but recent work paints a more complicated picture \citep{oo11}. It has been well-established that most long gamma-ray bursts (which may follow fallback accretion) are associated with broad-lines Type Ic supernovae \citep{wb06,hb11,mod11}, but such events are too distant to directly identify the exploding stars. Progenitor stars associated with standard \mbox{Type II-P} supernovae via pre-supernova imaging generally have masses $\lesssim17-20M_\odot$ \citep{sma09}, which is lower than the maximum mass expected for fiducial core-collapse scenarios \citep{heg03}. This might suggest that other types of supernovae, and maybe even BH formation, begin occurring in a lower mass range. Such a prospect is maybe not surprising given the historical difficulties in robustly producing supernovae via the neutrino mechanism in theoretical models \citep[although see][]{mue12}. On the other hand, \citet{smi11} argue the mass range of \mbox{Type II-P} progenitors are consistent with a substantial fraction ($\sim25-35\%$) of supernovae events being produced in binary systems, so the situation remains unresolved.

Gravitational waves (GWs) provide an independent probe to determine what processes are occurring deep within these massive stars \citep[for example, see the studies and reviews by][]{fry02,dim02a,dim02b,ott04,ott09,kot11}. The general picture we explore is as follows. Assuming that the fallback material forms a disk before reaching the NS, the NS accretes sufficient angular momentum that its spin parameter $\beta=T/|W|$ reaches a critical value $\beta_c$. 
Here $T$ is the rotational energy and $W$ is the gravitational binding energy.
Above $\beta_c$, non-axisymmetric instabilities occur and GWs are radiated (the exact value of $\beta_c$ depends on a number of factors, which are discussed in detail in \S \ref{sec:instability}). Since the star is quickly torqued down when $\beta>\beta_c$, and quickly spun up by accretion when $\beta<\beta_c$, the NS is forced into a state of marginal instability with $\beta\approx\beta_c$ while it continues to gain mass. The result is $\sim\unit[30-3000]{s}$ of high frequency ($\sim\unit[700-2400]{Hz}$) GW production until the NS becomes sufficiently massive to collapse to a BH. As we describe in more detail below, the detection of such a GW signal would be strong evidence that fallback accretion is occurring.

In \S \ref{sec:models}, we summarize the fallback model that is employed as well as our treatment of the non-axisymmetric shape of a quickly spinning NS. In \S \ref{sec:detection}, we provide a detailed discussion of the detection techniques using a excess cross-power search algorithm~\citep{stamp} to recover an example waveform. Given the inherent uncertainties in the messy process of fallback accretion, we consider this a much more representative assessment of the detectability than a naively optimistic matched-filtering approach, and we estimate that aLIGO can see such events out to $\approx\unit[17]{Mpc}$. In \S \ref{sec:background}, we discuss theoretical and observational constraints on the types of electromagnetic transients expected to be associated with such events. We conclude in \S \ref{sec:end} with a summary of our results, a discussion of future work, and some speculations about what astrophysics can be learned from these GW detections.


\section{The Fallback and Spinning NS Models}
\label{sec:models}

We begin by summarizing the main features of our models for fallback accretion and our treatment of spinning NSs. These semi-analytic models are clearly idealized, but they allow us to survey parameter space for this initial work. In future investigations, we plan to employ more realistic models based on numerically calculated progenitors.

\subsection{Fallback Accretion}

We consider fallback accretion onto a newly born NS as discussed by \citet{mac01} and \citet{zha08}. In particular, we focus on a progenitor with a $25M_\odot$ ZAMS mass with relatively inefficient semiconvective mixing \citep[referred to as model A in][]{mac01}. This star experiences mass loss during its lifetime and is reduced to $14.6M_\odot$ by the time core collapse occurs. \citet{mac01} input by hand a range of explosion energies from $2.6\times10^{50}-1.3\times10^{51}\ {\rm erg}$ and find a range of fallback masses from $0.24-3.7M_\odot$, where the two are inversely related. In all cases the remainder of the star that does not fall back is unbound. For energies greater than this, the entire envelope is ejected and a NS remnant is left.

To replicate the main features expected for fallback, we approximate the accretion rate as two power-laws \citep[as was done in][]{po11}. At early times it is relatively flat and scales as
\be
	\dot{M}_{\rm early} = \eta 10^{-3}t^{1/2}M_\odot\ {\rm s^{-1}},
	\label{eq:mdotearly}
\ee
where $\eta\approx0.1-10$ is a factor that accounts for different explosion energies (a smaller $\eta$ corresponds to a larger explosion energy), and $t$ is measured in seconds. This scaling of $\eta$ by two orders of magnitude corresponds to a change of the explosion energy from $2.6\times10^{50}-1.2\times10^{51}\ {\rm erg}$, which is merely a factor of five difference. This demonstrates just how sensitive the accretion rate is to the explosion energy. The late time accretion is roughly independent of the explosion energy and just depends on the mass of the progenitor at the onset of collapse. It is set to be
\be
	\dot{M}_{\rm late} = 50t^{-5/3}M_\odot\ {\rm s^{-1}}.
	\label{eq:mdotlate}
\ee
Interpolating these two expressions, we use
\be
	\dot{M} = \lp \dot{M}_{\rm early}^{-1} + \dot{M}_{\rm late}^{-1}  \rp^{-1}
	\label{eq:mdot}
\ee
for the accretion rate at any give time

For a NS with initial mass $M_0$, the time-dependent mass is
\be
	M(t) = M_0 +\int_0^{t}\dot{M}dt.
	\label{eq:mass}
\ee
Note that $M(t)$ corresponds to the total baryonic mass of the NS, but as discussed in \citet{lp01}, a non-negligible fraction of this mass becomes binding energy and is radiated away in the form of neutrinos. For a baryonic mass $M_{\rm baryon}$, the gravitational mass of a remnant with radius $R_{\rm grav}$ is
\be
	M_{\rm grav} = M_{\rm baryon} \lp 1+\frac{3}{5}\frac{GM_{\rm baryon}}{R_{\rm grav}c^2} \rp^{-1}.
\ee
Depending on the mass and radius, this can amount to a $\approx5-30\%$ correction. Given that we are just roughly approximating the accretion rate and that the conversion from a baryonic mass to a gravitational mass will occur in a time-dependent manner, we ignore this complication for this initial study.

\subsection{Spinning NS Model}

Besides increasing the NS mass, accretion also causes the NS to gain angular momentum. We assume that the infalling material roughly circularizes before reaching the NS surface. For this to occur, it must have specific angular momentum $j\gtrsim (GM_0R_0)^{1/2}\approx2\times10^{16}\ {\rm cm^2\ s^{-1}}$, where $R_0\approx 20\ {\rm km}$ is the nonrotating radius of the young NS. For material initially at a radius of $10^{10}\ {\rm cm}$, this corresponds to a rotational velocity of merely $\approx20\ {\rm km\ s^{-1}}$, so it appears quite likely that the NS accretes from a disk. The torque the NS experiences is therefore
\be
	N_{\rm acc} \approx \dot{M}(GMR_e)^{1/2}
	\label{eq:nacc}
\ee
where $R_e$ is the NS radius at the equator. We use $R_e$ to differentiate it from the nonrotating radius $R_0$ since in general $R_e\gtrsim R_0$ for a rotating body.

Equation (\ref{eq:nacc}) uses the accretion rate directly from fallback of the envelope, but more exactly this accretion rate should reflect the rate that mass is transferred through the disk. To explore whether this leads to a quantitative change of the accretion rate, we built one-zone, $\alpha$-disk models \citep[similar to][]{met08} using the angular momentum profiles of the massive, rotating progenitors of \citet{wh06}. Our general finding was that the disk reaches nearly steady state, where the accretion rate onto the star differs from the infall rate by no more than a factor of $\sim5$ (and this scales with the $\alpha$-viscosity, with a larger $\alpha$ resulting in higher accretion rates). We therefore consider the mediation of the disk to be degenerate with $\eta$ and use the direct infall rates as described above. In future studies using infall from numerical simulations, we plan to include these effects of a disk.

The rotation rate of the NS is measured in terms of the spin parameter $\beta\equiv T/|W|$. In the absence of an energy loss mechanism, the equilibrium shape of the spinning NS is characterized as an axisymmetric, Maclaurin spheroid. The relation between $\beta$ and the ellipticity $e$ for these figures is \citep{cha69}
\be
	\beta = \frac{3}{2e^2}\lb1-\frac{e(1-e^2)^{1/2}}{\sin^{-1}e} \rb-1
	\label{eq:beta}
\ee
where $e^2 = 1-(R_z/R_e)^2$ and $R_z$ is the vertical radius oriented along the spin axis. Although this result is for an incompressible fluid, it is also roughly valid for compressible configurations for our use here \citep{lai93}, since we never consider anything that is more than very mildly triaxial.

\begin{figure}
\epsscale{1.2}
\plotone{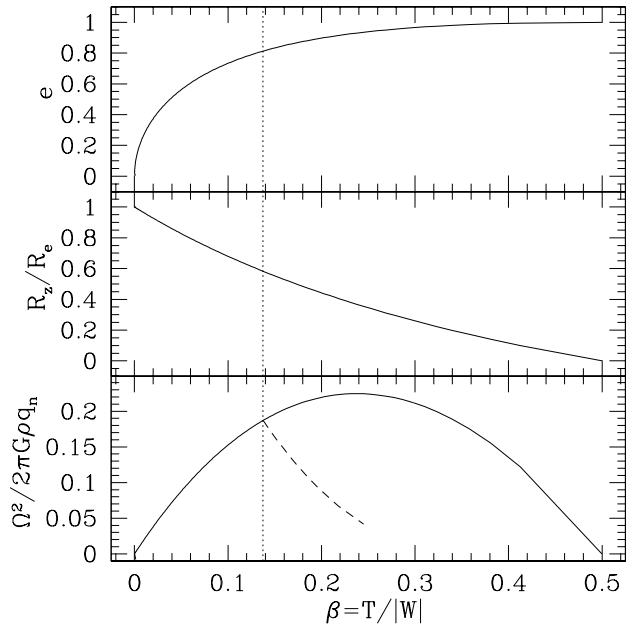}
\caption{General solutions for a Maclaurin spheroid given by equations (\ref{eq:beta}) and (\ref{eq:omega}). The vertical dotted line indicates the location of $\beta=\beta_{\rm sec}=0.14$. The dashed line in the bottom panel shows the ellipsoidal Jacobi solutions that branch off at $\beta_{\rm sec}$.}
\label{fig:ellipsoids}
\epsscale{1.0}
  \end{figure}

The spin of the NS is given by
\be
	\Omega^2 = \frac{2\pi G \bar{\rho}}{q_n}
	\lb \frac{(1-e^2)^{1/2}}{e^3}(3-2e^2)\sin^{-1}e - \frac{3(1-e^2)}{e^2} \rb,
	\label{eq:omega}
\ee
where $\bar{\rho}=3M/4\pi R_0^3$ is the average density and $q_n=(1-n/5)\kappa_n$ with $n$ as the polytropic index and $\kappa_n$ as a constant of order unity \citep[see Table 1 in][]{lai93}. The dimensionless solutions for Maclaurin spheroids given by equations (\ref{eq:beta}) and (\ref{eq:omega}) are summarized in Figure \ref{fig:ellipsoids}. Also shown in the bottom panel as a dashed line is the frequency dependence of the Jacobi ellipsoids, which are mentioned later. These must be found numerically, and the plotted solutions are from \citet{cha62}.

The mean radius of the rotating star $R=(R_zR_e^2)^{1/3}$ is given by
\be
	R = R_0 \lb \frac{\sin^{-1}e}{e}(1-e^2)^{1/6}(1-\beta)\rb^{-n/(3-n)},
\ee
so that
\be
	R_e = \frac{R}{(1-e^2)^{1/6}}.
	\label{eq:re}
\ee
This shows how the equatorial radius may be dramatically increased when the eccentricity is high, which in turn increases the specific angular momentum of accreted material.

\begin{figure}
\epsscale{1.2}
\plotone{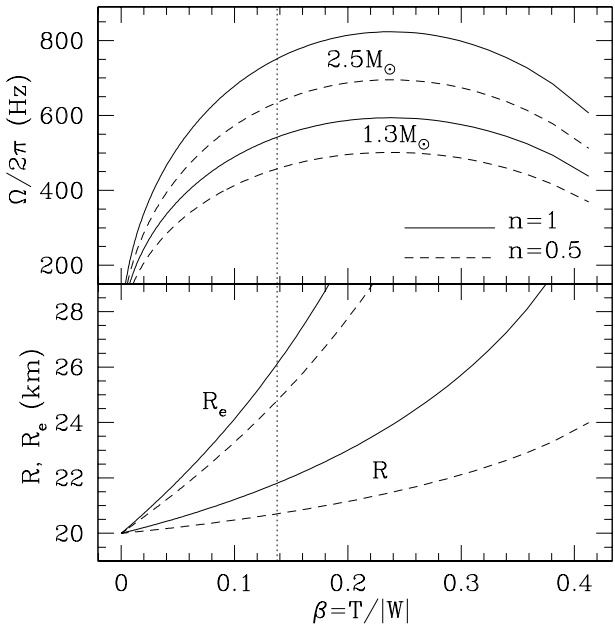}
\caption{Spin frequency $\Omega$, average radius $R$, and equatorial radius $R_e$ as a function of the spin parameter $\beta$, comparing $n=0.5$ (dashed line) and $n=1$ (solid line). The NS has either $M=1.3M_\odot$ or $M=2.5M_\odot$ (as labeled in the upper panel), but in either case $R_0=20\ {\rm km}$. The critical $\beta_{\rm sec}=0.14$ for secular instability is plotted as a vertical, dotted line.}
\label{fig:ns}
\epsscale{1.0}
\end{figure}

In Figure \ref{fig:ns} we plot sequences of Maclaurin spheroids for a NS with $M=1.3M_\odot$ or $M=2.5M_\odot$ (as labeled in the upper panel) and $R_0=20\ {\rm km}$, comparing polytropic indices of $n=0.5$ (dashed lines) with $n=1$ (solid lines). This brackets the range of reasonable values for NSs. In the top panel we plot the spin frequency as a function of $\beta$. As the mass increases from $1.3M_\odot$ to $2.5M_\odot$, the corresponding spin frequency increases at fixed $\beta$. Similarly, the more centrally concentrated $n=1$ polytrope generally exhibits a higher spin frequency than the $n=0.5$ polytrope. In the bottom panel we plot both the average radius $R$ and the equatorial radius $R_e$. The equatorial radius is strictly increasing as a function of $\beta$, demonstrating how the shape is becoming increasingly like a flattened pancake. Changes in the average radius are more modest.

One issue that deserves mention is that for $n=1$, solid-body rotating models have equatorial velocities that exceed Keplerian at merely $\beta\approx0.08$. This means that such models are not self-consistent, but they can be stabilized against equatorial mass loss with a small amount of differential rotation \citep{ls95}. For now we ignore this detail, but also consider $n=0.5$ polytropes that are at least stable against mass shedding up to $\beta\approx 0.15$.

\subsection{Instability and GW Production}
\label{sec:instability}

Although these equilibrium Maclaurin spheroids can be found for all $\beta$ plotted in Figure \ref{fig:ns}, such configurations are not in general stable, and liable to break their symmetry and transition to a nonaxisymmetric shape. When $\beta$ exceeds a some critical $\beta_c$ and instability sets in, the resulting triaxial shape causes the production of GWs. This in turn limits $\beta$ from any further growth. On the other hand, if $\beta$ is forced below $\beta_c$, GW production ceases, and the NS spins up again. Due to these competing effects, we expect the NS spin parameter to reach a saturated state where $\beta\approx\beta_c$. The NS then continues to accrete as it produces GWs at a rate that maintain this balance.

There are a number of potential values for $\beta_c$ that could be considered. For example, it has been well-established that a dynamical bar-mode instability sets in when $\beta>\beta_{\rm dyn}=0.27$. This leads to mass shedding and spindown back to a stable state \citep{shi00}. At lower values of $\beta$, but still greater than $\beta_{\rm sec}=0.14$, the possible solution for the spinning NS can also be triaxial and given by a Jacobi ellipsoid (as plotted in Figure \ref{fig:ellipsoids}) or Dedekind ellipsoid (not plotted since formally these figures are not rotating). These represent lower-energy configurations, but the NS can only trigger these so-called secular instabilities and transition from a spheroidal Maclaurin solution to these solutions if acted upon by some sort of dissipative process. Dynamical shear instabilities may also operate for $\beta\gtrsim0.01$ if differential rotation is present \citep{cen01,shi03,wat05,ot06,cor10}. Since the conditions required for these low $\beta$ instabilities and their associated GW signature are more complicated, we focus on the secular instability at $\beta_{\rm sec}$ for the majority of this work.

Potentially destabilizing mechanisms that may trigger instability for $\beta>\beta_{\rm sec}$, and that have been well-studied in the literature, are viscosity and gravitational radiation reaction \citep{cha70,fs78,lai01,gk11}. A rough estimate for the  growth time due to destabilization from gravitational radiation is
\be
	\tau_{\rm gw} \approx 2\times10^{-5}M_{1.3}^{-3}R_{20}^4(\beta-\beta_{\rm sec})^{-5}\ {\rm s},
	\label{eq:taugw}
\ee
which is roughly independent of $n$ \citep{ls95}. This destabilizes the Dedekind-mode, which corresponds to a highly differentially rotating, but stationary figure with a rotation pattern that gives rise to a bar-mode like oscillation. In comparison, typical timescales for accretion are $\tau_{\rm acc}\approx M/\dot{M}\sim 10^2-10^4\ {\rm s}$. For gravitational radiation reaction to be effective, it must therefore act in a timescale $\tau_{\rm gw}\lesssim \tau_{\rm acc}$, which implies $\beta\gtrsim 0.16-0.18$. Although not far above $\beta_{\rm sec}=0.14$, it is possible that other destabilizing processes can occur first as the NS spin is increasing. For example, the accretion itself or viscous processes near the surface of the NS may be extremely dissipative. For these reasons we consider it reasonable to assume that $\beta_c\approx \beta_{\rm sec}$, but whether there are small variation in $\beta_c$ for the particular case of fallback accretion deserves further study in the future.

The next important question is what kind of rotating figure is present during this saturated state. In general, there are a wide range of spheroidal and ellipsoidal figures available, but for the moment we concentrate on the Dedekind and Jacobi ellipsoids. Although these belong to the much larger class of Riemann-S ellipsoids, the general features of our argument still apply. The Dedekind-mode is destabilized by GWs because of its stationary properties. This allows it to have an equilibrium state without GW production, and at the same time it exhibits high internal differential rotation to conserve its circulation. In a sense, the stable properties of the Dedekind-mode are somewhat unique to the effect of GWs, which only care about the time-varying quadrupole moment. In contrast the Jacobi-mode is destabilized for any viscosity. Furthermore, its rigidly rotating interior (with perhaps a small amount of differential rotation to prevent mass shedding) seems more representative of the interior of a NS, which is subject to a wide range of potential viscosity mechanisms as well as magnetic fields that resist any differential rotation. Given the above arguments, the spinning NS is best approximated as a Jacobi-like ellipsoid that is roughly rigidly rotating and, as we will show below, is actually very close to just being spheroidal.

It should be noted that our model is decidedly different than that employed by \citet{cm09} to follow NS spin evolution following a magnetar-powered long gamma-ray burst. They focus on spindown by gravitational waves and dipole radiation and include no fallback accretion, nor any magnetic or viscous effects on the internal fluid motions of the star. Because of this, their spin evolution asymptotes to a highly differentially rotating Dedekind ellipsoid instead.

In this saturated state the NS emits GWs with an associated energy loss of
\be
	\dot{E}_{\rm gw} = -\frac{32G\Omega^6}{5c^5} (I_{11}-I_{22})^2,
	\label{eq:egw}
\ee
where $I_{ii} = \kappa_n Ma_i^2/5$ are the components of the star's quadrupole, and $a_1$ and $a_2$ are the 
main axis of the triaxial figure. Angular momentum is removed at a rate
\be
	N_{\rm gw} = \dot{E}_{\rm gw}/\Omega.
	\label{eq:ngw}
\ee
The corresponding GW strain amplitude as measured on Earth is \citep{bra98}
\be
	h_0 = \frac{2G\Omega^2}{c^4D}(I_{11}-I_{22}),
	\label{eq:h0}
\ee
where $D$ is the distance to the source.

Given the above discussion, we summarize our procedure for estimating the GW emission from fallback accretion as follows.

(1) Accretion increases $\beta$ on a timescale $\sim10-50\ {\rm s}$ until at a value $\beta\approx\beta_c$,  GW production saturates the spin parameter. If the NS is destabilized due to gravitational radiation reaction, then $\beta_c\approx0.16-0.18$. But given the potential for viscous or accretion driven destabilization, we consider it reasonable to set $\beta_c\approx\beta_{\rm sec}$.

(2) This $\beta_c$ implies an ellipticity $e$, spin frequency $\Omega$, and equatorial radius $R_e$ via equations (\ref{eq:beta}), (\ref{eq:omega}), and (\ref{eq:re}) respectively.

(3) For GW emission to maintain $\beta\approx\beta_c$, it must remove angular momentum at a rate that roughly balances the accretion torque, so that $N_{\rm gw}+ N_{\rm acc}\approx0$.

(4) Using this equilibrium condition, we can solve for $I_{11}-I_{22}$ using equations (\ref{eq:egw}) and (\ref{eq:ngw}), which can then be substituted into equation (\ref{eq:h0}) to find $h_0$.

\subsection{Estimates and Time-Dependent Evolutions}

To better demonstrate how our GW calculations will proceed, it is useful to present an example implementation. Consider accretion onto an $M_0=1.3M_\odot$, \mbox{$R_0=20\ {\rm km}$} NS. Setting $\beta\approx \beta_c$ and using $\beta_c=\beta_{\rm sec}$, implies that the ellipticity and spin in the saturated state are $e=0.813$ and $\Omega/2\pi=542\ {\rm Hz}$ ($n=1$) or $\Omega/2\pi=457\ {\rm Hz}$ ($n=0.5$). For $n=1$, the mean radius is $R= 1.09R_0$ and in turn the equatorial radius is $R_e=1.31R_0$. For $n=0.5$, $R=1.035R_0$ and $R_e=1.24R_0$.  These values match what is plotted in Figure \ref{fig:ns} at $\beta=\beta_{\rm sec}$ (along the vertical dotted line), and are consistent with the general intuition that a more centrally concentrated mass distribution (i.e., higher $n$) is more strongly effected by rotation at a given $\beta$.

\begin{figure}
\epsscale{1.2}
\plotone{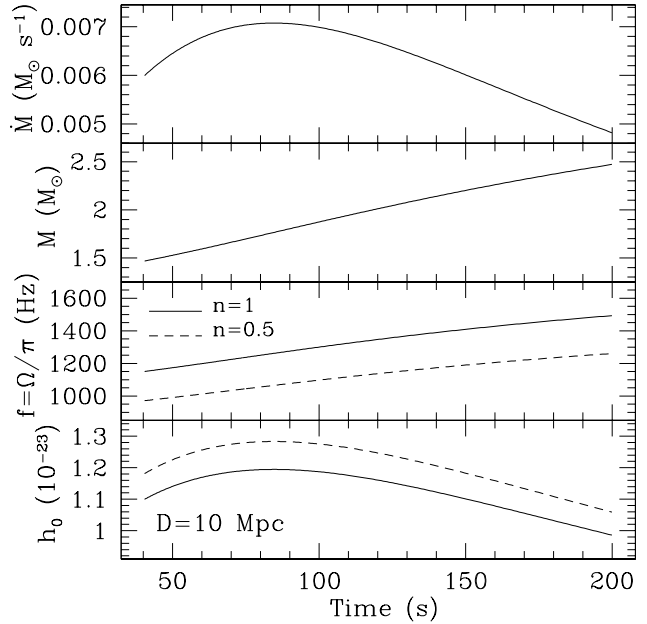}
\caption{Example frequency, mass, and strain amplitude evolution using the scheme described in these notes. The NS has $M_0=1.3M_\odot$ and $R=20\ {\rm km}$, with both $n=1$ (solid curves) and $n=0.5$ (dashed curves), and $\eta=1.0$ for setting $\dot{M}$. It is assumed that the NS accretes for $\unit[40]{s}$ during which $\beta<\beta_{\rm crit}$ and no angular momentum is lost. We then set $\beta=\beta_{\rm crit}=0.14$ and follow the evolution up until $M\approx2.5M_\odot$, at which point the NS would collapse to become a BH.}
\label{fig:strain}
\epsscale{1.0}
  \end{figure}

In Figure \ref{fig:strain} we present  example time evolutions of accreting NSs in the saturated state. The top two panels show the accretion rate from equation (\ref{eq:mdot}) using $\eta=1$ and the total mass given an initial mass of $M_0=1.3M_\odot$ and using equation (\ref{eq:mass}). It is assumed that the NS accretes for $40\ {\rm s}$ before $\beta=\beta_{\rm crit}$, and thus earlier times are not shown in this plot. We could potentially attempt to set the timescale at which the saturated state begins self-consistently with the early time accretion rate. The problem is that when this occurs will depend on the initial spin of the core, which is difficult to estimate. We therefore leave the initial time for saturation as a free parameter and consider any time from $\sim10-50\ {\rm s}$ to be reasonable. As described previously, a given mass and value of $\beta_c$ is sufficient to solve for the spin rate (or GW frequency), which is plotted in the third panel of \mbox{Figure \ref{fig:strain}.} This depends on the density distribution of the star, and thus we plot $f$ for both $n=1$ (solid lines) and $n=0$ (dashed lines) polytropes.

When the GW torque is in equilibrium with the accretion torque ($N_{\rm gw}+N_{\rm acc}\approx0$), the quadrupole moment $Q=I_{11}-I_{22}$ is
\be
	Q = \lb \frac{5}{32}\frac{(GMR_e)^{1/2}\dot{M}c^5}{G\Omega^5} \rb^{1/2}.
\ee
For $\dot{M}=10^{-3}M_\odot\ {\rm s^{-1}}$, we estimate a poloidal ellipticity $\epsilon\equiv Q/MR_0^2=7.0\times10^{-3}$ ($n=1$) or $\epsilon=1.1\times10^{-2}$ ($n=0.5$). This demonstrates that the NSs we are considering here are only deviating from sphericity by a small amount. This gives us some confidence that accretion will proceed unhindered by the triaxial shape of the NS. (In contrast, it is unlikely that accretion can occur onto a NS that is spinning fast enough for dynamical instabilities and has an extreme bar-mode shape.) It should be noted though that $Q/MR_0^2\gtrsim10^{-5}$ exceeds the maximum strain that a neutron crust can support \citep{sha77}. This crustal physics is not included in our calculation, and a future, more detailed investigation should include the time-dependent effects of crust formation.

Substituting this $Q$ into the strain in equation (\ref{eq:h0}), and using a distance of $D=10\ {\rm Mpc}$, gives $h_0 = 4.4\times10^{-24}$  ($n=1$) and $4.8\times10^{-24}$ ($n=0.5$). We also plot the time-dependent strain amplitude in the bottom panel of Figure \ref{fig:strain}. This is similar, but slightly larger than these estimates since the NS and its spin frequency are increasing with time even as $\beta$ remains fixed.



\section{Detectability of simulated waveforms}
\label{sec:detection}
In this section we survey upcoming GW interferometers and assess the detectability of GWs from a NS subject to fallback accretion. Focusing on realistic excess cross-power searches, we show that second-generation detectors such as Advanced LIGO (aLIGO) may be able probe accretion-powered NS models out to astrophysically interesting distances: $\approx\unit[17]{Mpc}$.

\subsection{Gravitational-wave interferometers}
Recent years have seen the development of a worldwide network of GW interferometers.
The initial LIGO experiment~\citep{iligo} achieved design strain sensitivity: $\approx\unit[2\times10^{-23}]{Hz^{1/2}}$ in the most sensitive frequency range of $\unit[100\text{-}200]{Hz}$.
The next generation of interferometers such as Advanced LIGO (aLIGO)~\citep{aligo}, Advanced Virgo~\citep{virgo}, and \mbox{KAGRA~\citep{kagra}} are expected to begin taking data starting in $\approx2015$, while the collection of data from GEO~\citep{geo} is ongoing.
The upgraded aLIGO/Virgo experiments are expected to achieve a factor of ten higher better strain sensitivity than initial LIGO/Virgo.
In our simulation study below we show that aLIGO can probe fallback accretion signals to objects as distant as the Virgo Cluster.

\subsection{Waveform simulation}
In order to assess the detectability of fallback accretion events, we inject simulated waveforms into Monte Carlo noise and determine the distance at which signals can be distinguished from noise.
Far from the source, we can write the metric perturbation as
\be
	h_+(t) = h_0(1+\cos^2\iota)\cos\Phi(t)
\ee
and
\be
	h_\times(t) = 2h_0 \cos \iota \sin\Phi(t)	
\ee
where $h_0$ is calculated from equation (\ref{eq:h0}),
\be
	\Phi(t) = \int_0^t 2\pi f(t') dt', 
    \label{eq:Phi}
\ee
is the time-dependent phase, and $\iota$ is the inclination angle, (which we later set to zero by assuming that the source is face on)\footnote{Following~\cite{prix}, we work in a right-handed basis $\{\hat\mu,\hat\nu,-\hat{n}\}$ where $\hat{n}$ points from the detector to the source, and ${\hat\mu, \hat\nu}$ are aligned the principal polarization axes of the GW.}.
We calculate $h_+(t),h_\times(t)$ by numerically evaluating the integral in equation~\ref{eq:Phi}.

\subsection{Injection recovery}
We envision an excess cross-power search \citep[see][]{stamp,stamp_glitch} for long-lived GWs associated with a well-localized electromagnetic counterpart, so that the direction and time of the signal are constrained, (though more computationally expensive ``all-sky'' searches are possible as well).
We consider a simple network consisting of two $\unit[4]{km}$ initial LIGO observatories operating at design sensitivity; one in Hanford (H1) and one in Livingston (L1).

We assume that the GW frequency of the signal is between $\unit[700\text{-}2400]{Hz}$, which is true for the models considered here (see Table~\ref{tab:distances}).
We assume that the electromagnetic trigger constrains the on-source region (during which the signal may be present) to $\unit[1000]{s}$ in duration. Such an on source region is similar to the timescale over which the start of a supernova can be identified from shock breakout of a compact progenitor \citep{ns10}, and considerably longer than the typical timescale of a long gamma-ray burst. On the other hand, if the supernova progenitor is more extended (like a red supergiant), then the on source region could be $\sim3-10$ times larger \citep{sch08}.
We explore how sensitivity varies with the on-source duration below.
We assume that electromagnetic measurements constrain the direction of the source to better than $0.17^\circ$, inside which at least 90\% of the H1-L1 point-spread function is contained for a source with $f\leq\unit[2400]{Hz}$.
(If the source cannot be this well localized, the sky can be tiled in order to search over multiple directions.)

\begin{figure*}
  \begin{tabular}{cc}
    \psfig{file=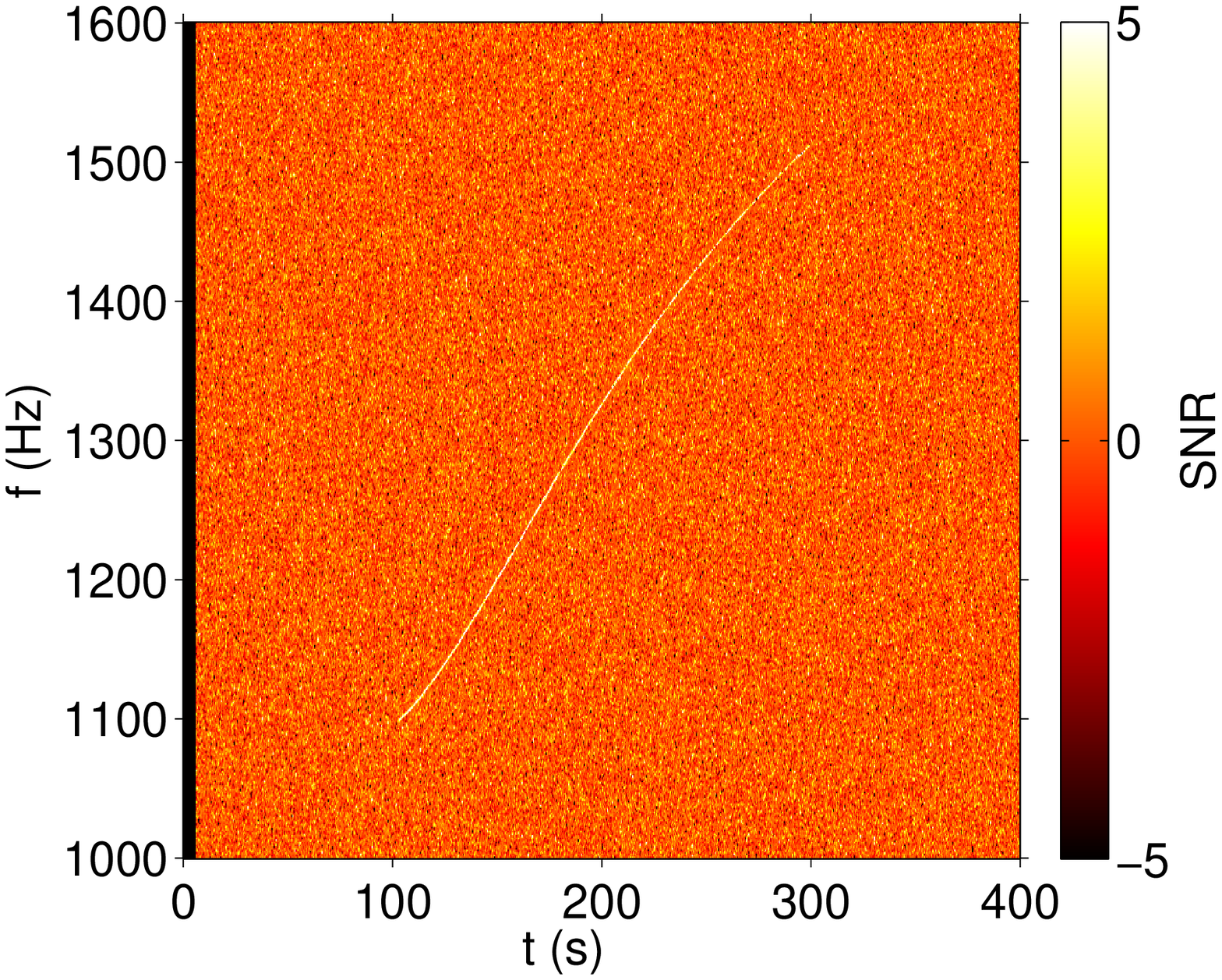, height=2.5in} &
    \psfig{file=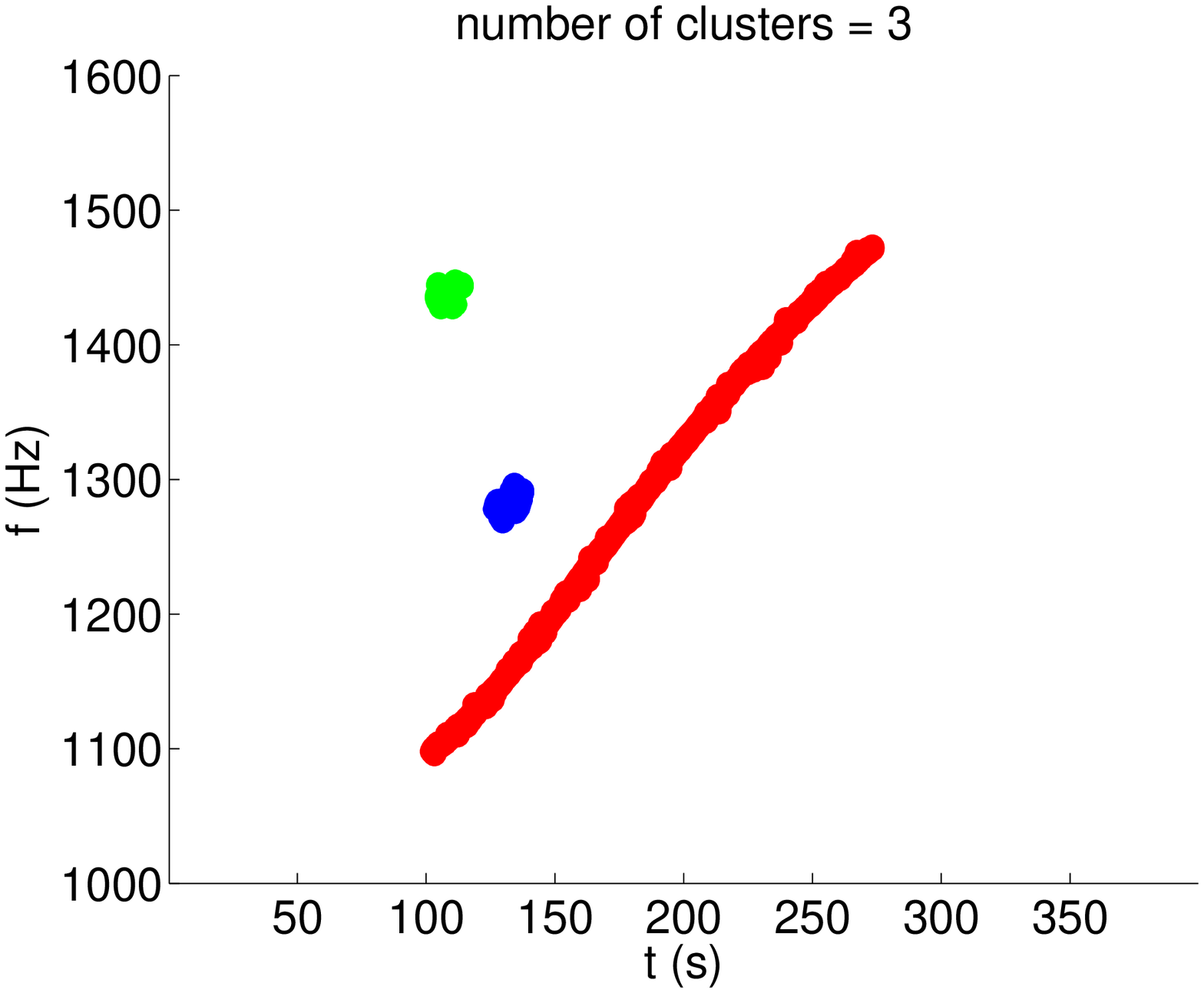, height=2.5in} \\
    \psfig{file=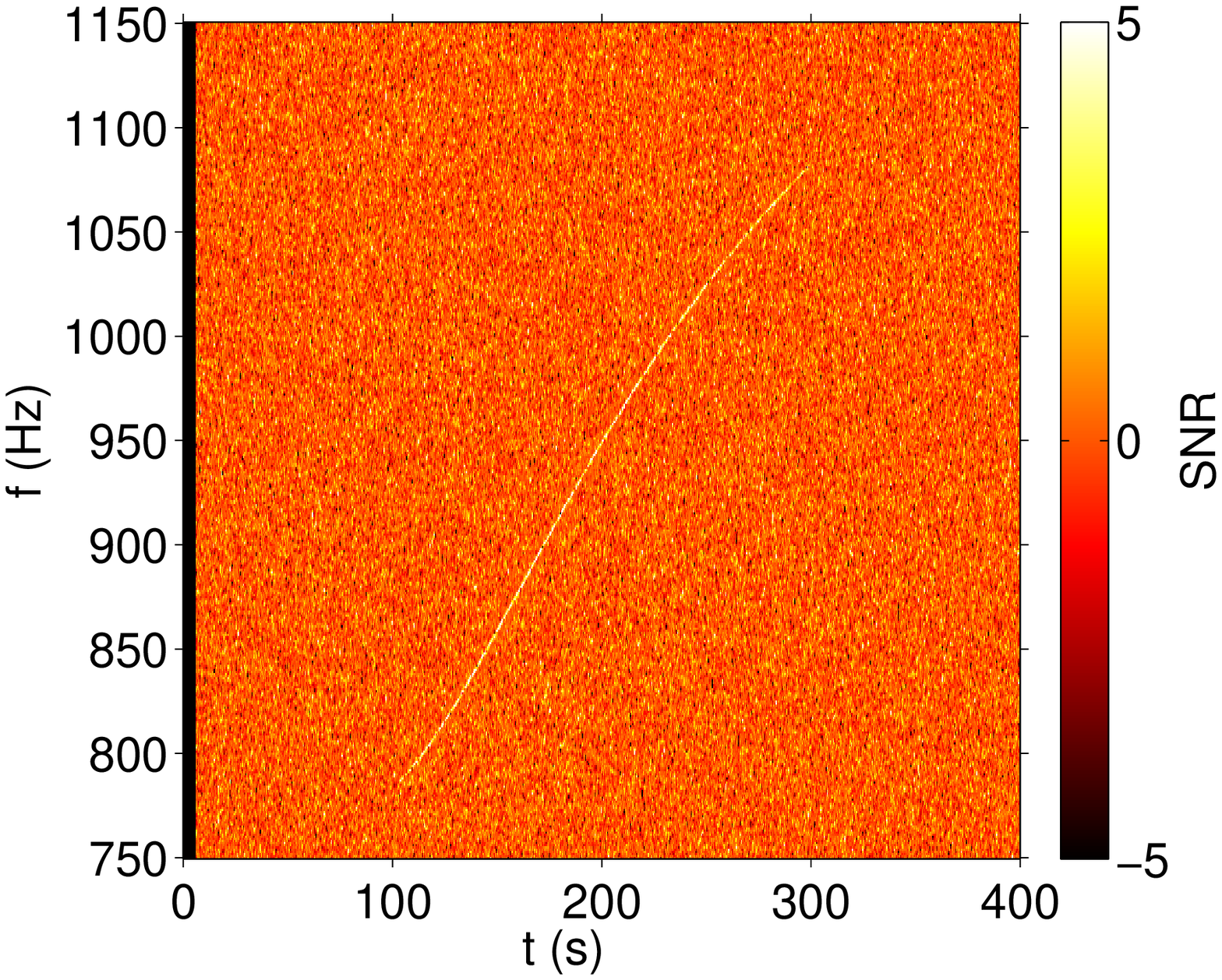, height=2.5in} &
    \psfig{file=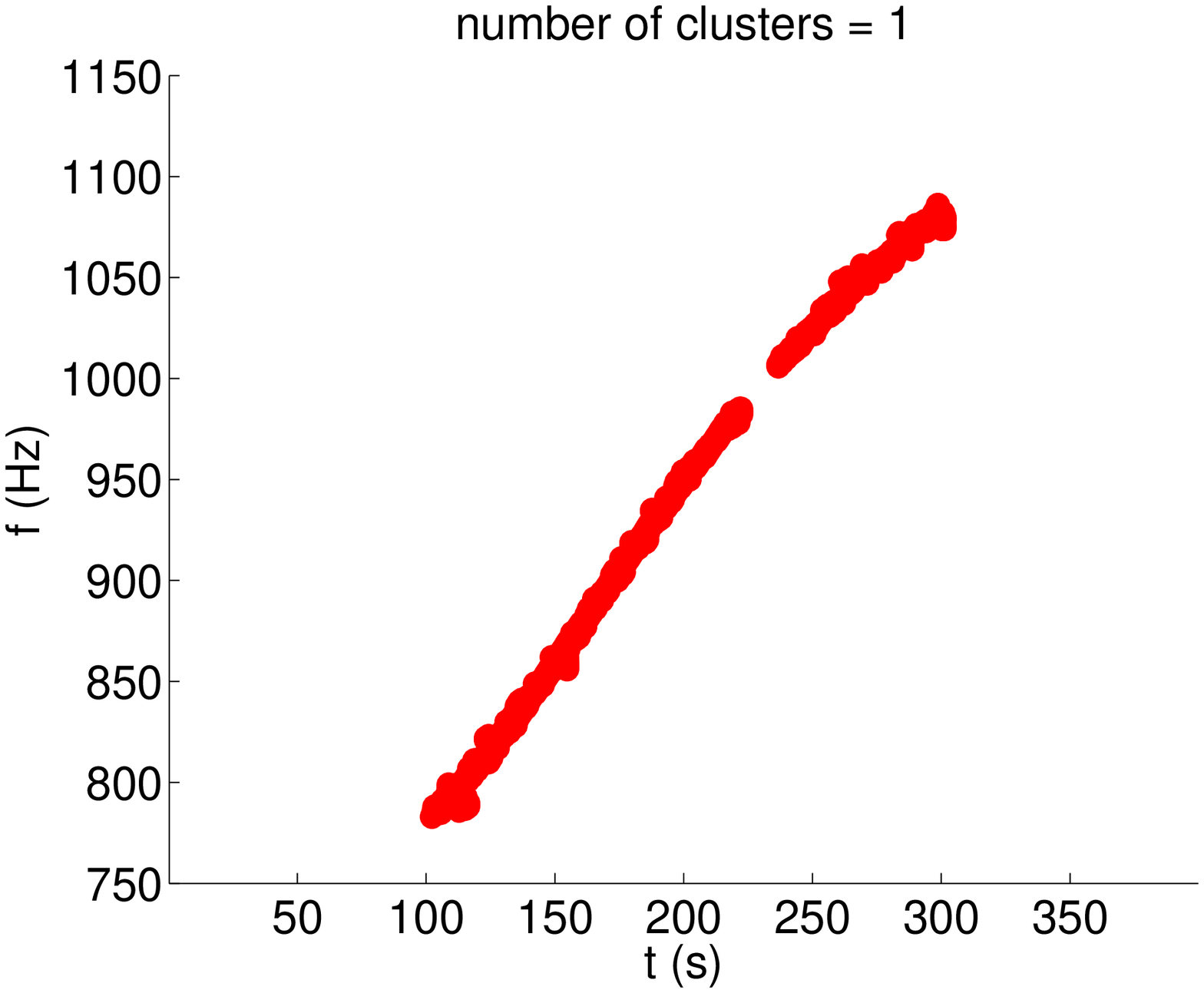, height=2.5in} \\
  \end{tabular}
  \caption{
    Recovered signals injected into aLIGO Monte Carlo noise: $M=1.3M_\odot$, $R=\unit[20]{km}$, $\eta=1$, $D=\unit[5]{Mpc}$ (top) and $M=1.3M_\odot$, $R=\unit[25]{km}$, $\eta=1$, $D=\unit[10]{Mpc}$ (bottom).
    Left: a spectrograms of $\text{SNR}(t;f)$.
    We have cropped the spectrogram from its full size so that the signal is visible by eye.
    Right: candidate clusters identified by the track-search algorithm.
    (The small green and blue clusters in the top-right plot are noise fluctuations, while the long, thin, red track is the reconstructed GW signal.)
    The cluster SNR, $\text{SNR}(c)=168$, $111$ (top, bottom), is well above the 0.1\% false alarm probability threshold of 23.
  }
  \label{fig:injection}
\end{figure*}

Within this $\unit[1000]{s}\times\unit[1700]{Hz}$ on-source region, and following~\cite{stamp}, we create a spectrogram of signal-to-noise ratio $\text{SNR}(t;f)$, which is proportional to the cross-correlation of the H1 and L1 strain data.
Here $f$ refers to the frequency bin in a discrete Fourier transform centered on time $t$.
We use $\unit[1]{s}$-long, 50\%-overlapping, Hann-windowed data segments, which yield spectrograms with a resolution of $\unit[0.5]{s}\times\unit[1]{Hz}$.
Note that
\begin{equation}
  \langle\text{SNR}(t;f)\rangle = h_0^2 \left( 1-\cos^2\iota \right) ,
\end{equation}
and that noise fluctuations can be both positive and negative.
A formal derivation of $\text{SNR}(t;f)$ can be found in~\cite{stamp}.
An example of a $\text{SNR}(t;f)$ spectrogram can be seen in the two left panels of Figure~\ref{fig:injection}.

The problem of detection is to identify a track of positive-valued pixels in the presence of noise.
We employ a track-search algorithm~\citep{burstegard}, which looks for clusters of positive pixels\footnote{For other examples of clustering algorithms used in GW data analysis, see~\citet{burstcluster} and \citet{raffai}.}.
The pixels are combined to determine the SNR for the entire cluster denoted $\text{SNR}(c)$.
Note that $\text{SNR}(c)$ is distinct from $\text{SNR}(t;f)$, which is the SNR associated with individual pixels in a spectrogram.
We apply the track-search algorithm to Monte Carlo Gaussian noise in order to determine the threshold for an event with false alarm probability 0.1\%.
We find this threshold to be $\text{SNR}(c)=23$.

\begin{deluxetable*}{c c c c c c c}
  \tablecolumns{5} \tablewidth{0pt}
  \tablecaption{Fallback accretion models with associated aLIGO Detection Distances\tablenotemark{a}}
  \tablehead{
    \colhead{$M_\text{max}$ ($M_\odot$) } & 
    \colhead{$R_0$ (km)} & 
    \colhead{$\eta$} & 
    \colhead{$f_\text{min}$-$f_\text{max}$ (Hz)} & 
    \colhead{$t_\text{dur}$ (s)} & 
    \colhead{$D_\text{burst}$\tablenotemark{b} (Mpc)} & 
    \colhead{$D_\text{mf}$\tablenotemark{c} (Mpc)} }
  \startdata
  2.5 & 20 & 1 & 1098-1512 & 208 & 9.5 & 125 \\
  2.9 & 20 & 1 & 1098-1655 & 354 & 10.0 & 135 \\
  2.5 & 15 & 1 & 1690-2328 & 208 & 5.0 & 31 \\
  2.5 & 25 & 1 & 785-1082 & 208 & 17.0 & 195 \\
  2.5 & 20 & 10 & 1174-1528 & 37 & 12.5 & 115 \\
  2.5 & 20 & 0.5 & 1093-1508 & 385 & 8.5 & 120 \\
  2.5 & 20 & 0.3 & 1091-1510 & 641 & 7.5 & 145 \\
  2.5 & 20 & 0.1 & 1089-1510 & 3100 & $<7.5\tablenotemark{d}$ & ---
   \enddata
   \tablenotetext{a}{Distances are calculated for a false alarm probability $<0.1\%$ and a false dismissal probability $<50\%$ using Monte Carlo aLIGO noise (high-power, zero-detuning~\citep{aligo_noise}). We assume an initial NS mass of $1.3 M_\odot$. Note, for comparison, that the Virgo Cluster is $\approx\unit[16.5]{Mpc}$ away from Earth, while the Andromeda Galaxy is $\approx\unit[0.8]{Mpc}$ away. Sources are assumed to be optimally oriented.}
   \tablenotetext{b}{Detection distance for a realistic excess cross-power search.}
   \tablenotetext{c}{Detection distance for a highly idealized matched-filtering search.}
   \tablenotetext{d}{Due to its length, more work beyond our present scope is required to carefully study this $\unit[3100]{s}$ signal.  We note that the detection distance is likely to be less than a signal with identical parameters except for a larger $\eta$. When $\eta$ is decreased, the power is spread out more diffusely over a longer duration, making the signal harder to detect.}
   \label{tab:distances}
\end{deluxetable*}

Real interferometer strain noise is only approximately Gaussian and includes noise bursts (glitches) and other non-stationary effects.
As a rule of thumb, glitches have a tendency to push the required detection threshold higher than it would be in Gaussian noise.
However, it was shown in~\cite{stamp_glitch} that initial LIGO cross-correlated data can be cleaned to the point where it is essentially indistinguishable from ideal Gaussian noise in the context of a search for long ${\cal O}(\gtrsim\unit[10]{s})$ narrowband GW transients.
This is especially true in the shot-noise regime at high frequencies $f\gtrsim\unit[500]{Hz}$ where our signal resides.
Thus, we expect that the threshold of\mbox{ $\text{SNR}(c)=23$} is representative of what can be achieved with realistic interferometer data.

Next, we inject simulated signals into the Monte Carlo noise in order to determine the amplitude above which 50\% of the signals are recovered above threshold.
The signals are injected assuming an optimally oriented source ($\iota=0$) and in the direction that maximizes the interferometer network sensitivity.
(The sky-averaged sensitivity is $\approx2.5$ less than the optimal sensitivity.)
Plots illustrating the recovery of injected signals are presented in the two right panels of Figure~\ref{fig:injection}.
By varying the distance to the source, we can determine the distance at which we can observe a signal with false alarm probability  $\leq0.1\%$ and false dismissal probability 50\%.
This ``detection distance'' denoted $D_\text{burst}$ is given in Table~\ref{tab:distances} for different combinations of parameters.
We also include the corresponding distance that can be obtained with a highly idealized matched filtering analysis $D_\text{mf}$. We return to the topic of matched filtering below.

We find that, depending on the source parameters, aLIGO can observe signals out to distances of \mbox{$\approx\unit[5-17]{Mpc}$}.
Thus, aLIGO will be able to probe fallback-powered GW emission from NSs in nearby galaxies such as Andromeda ($\unit[0.8]{Mpc}$) and perhaps as far away as the Virgo Cluster ($\unit[16.5]{Mpc}$).
These distance estimates are obtained using an existing pattern recognition search algorithm.
It is possible that the sensitivity can be improved by developing a specially tailored search algorithm.

Until now we have assumed that the GW signal can be confined to a $\unit[1000]{s}$ on-source region.
In order to determine how this assumption affects the search sensitivity, we also consider a longer $\unit[3000]{s}$ on-source region.
We find that the 0.1\% false alarm probability threshold grows from $\text{SNR}(c)=23$ to $\text{SNR}=25$.
Repeating the above injection study, we find that the detection distance for a $\unit[3000]{s}$ window is about 3\% smaller than for a $\unit[1000]{s}$ window---e.g., the detection distance for the $R=\unit[25]{km}$ waveform shown in Figure~\ref{fig:injection} changes from $\unit[17.0]{Mpc}$ to $\unit[16.5]{Mpc}$.
Thus, the falloff in sensitivity is modest as the on-source region is increased in size, which in turn suggests that the analysis we have sketched out can be performed for a wide range of electromagnetic triggers.
In the next section, we investigate the best theoretically possible (though probably unrealistic) sensitivity that can be achieved with matched filtering.

\subsection{Matched filtering estimates}
For GW signals of a known form, the theoretically optimal search strategy is matched filtering \citep[see, e.g.,][]{s6lowmass,s5cw_allsky,owen_sathya}.
The fallback accretion powered signals we consider here, however, will prove challenging to target with a matched filtering search.
Such a search must include a template bank that spans the space of possible signals.
An incomplete template bank can result in faulty upper limits if the true signal falls outside the template space.
In order to generate a complete template bank, we require firm knowledge of the details of the waveform's phase evolution.
In presenting the above model, we have not aspired to this degree of accuracy.
Moreover, even if a complete and accurate model could be written down, there may be computational challenges associated with performing the search, especially for long signals with many parameters.

Nonetheless, it is useful to compare the detection distances calculated using the excess-power technique to estimates for what can be achieved with matched filtering as this places an upper limit on the detection distance that can be achieved through improvements to the data-analysis scheme.
Rather than create a bank of templates to span the space of our signal, we assume that every signal parameter is known including: the start time of the signal, the initial phase, the initial mass $M_0$, the accretion parameter $\eta$, the polytrope index, the direction, and the critical value of $\beta$.
As before, we assume optimal orientation of the source, optimal orientation of the detector network, a false-alarm probability of 0.1\% and a false dismissal probability of 50\%.
This highly idealized analysis allows us to use just a single template.
We find that highly idealized matched filtering allows us to extend the detection distance by a factor of $6$ to $19$.
This factor varies depending in part on the efficiency of the pattern recognition algorithm for different signal types.

\subsection{Other Gravitational-Wave Signatures}
Besides the GWs from fallback, there may be other GW signals associated with supernovae and the creation of a BH. A variety of mechanisms have been proposed for GW emission from supernovae, including the acoustic mechanism, rotational instabilities, rotating collapse and bounce, convection and standing accretion shock instabilities \citep[see][and references therein]{ott09}.
Competing models predict GW fluence that spans six orders of magnitude ($2$ to $\unit[2\times10^6]{erg \ cm^{-2}}$ for a galactic source at $\unit[1]{kpc}$), but at least one model (the acoustic mechanism) predicts strain amplitude high enough for aLIGO to observe in Andromeda~\citep{gwnu,ott06}. 
It is therefore possible that that the fallback GWs we describe here could potentially be present from the same event that produces a short GW burst in Andromeda.

In addition to the precursor signal from core collapse, it is interesting to consider the possibility of a postcursor signal from the ringing black hole eventually formed through accretion~\citep{fry02,ott11}.
There are significant uncertainties surrounding the strain amplitude of the black hole ringdown signal following accretion-induced collapse, but the signal morphology is relatively well constrained~\citep{fry02}.
For a newly formed black hole with mass $2.5 M_\odot$, we expect a damped sinusoid signal with frequency $f\approx\unit[6.5]{kHz}$ and quality factor $Q\approx3$~\citep{s4ringdown}.
The strain amplitude depends on the fraction $\epsilon$ of energy radiated in GWs and the mass $\mu$ falling into the black hole, thereby exciting the quasinormal modes~\citep{fry02}.
If we optimistically assume $\epsilon=0.01$ and $\mu=0.1 M_\odot$, the strain amplitude is $h_0\approx6\times10^{-22}$, which implies a matched filtering detection distance of $\unit[0.15]{Mpc}$ (assuming an optimally oriented source, requiring a false alarm probability = 0.1\%, and a false dismissal probability = 50\%).
Thus the possibility of observing a ringdown signal is probably limited to sources in our own galaxy.


\section{Electromagnetic Counterparts and Rates}
\label{sec:background}

Given that fallback accretion onto a NS can be observed with aLIGO out to a distance of $\approx\unit[17]{Mpc}$, it is worth exploring the broader context of what kind of events could be associated with potential GW detections.  It has been well-established in the case of compact mergers that joint GW and electromagnetic detections provide the greatest opportunity for learning about these systems. This should likewise be true for core-collapse supernovae. In the following discussion we draw upon recent theoretical and observational progress in understanding core-collapse supernovae to address potential electromagnetic counterparts and estimate rates.

To get fallback accretion as needed for GW production requires a few key ingredients: (1) sufficient angular momentum for disk formation, (2) a sufficiently low explosion energy to not unbind the entire envelope, and (3) a progenitor that is sufficiently massive and compact for eventual BH formation. Disk formation within collapsing stars has traditionally been the focus of studies trying to produce long gamma-ray bursts. For example, a large part of the work by \citet{mac01}, which motivated the fallback model we employed, was presented to study the hyperaccretion that would occur {\it following} BH formation. Since we too require rotation to spin up the NS and generate GWs, this may mean that hyperaccretion and jet formation may follow. But this is not certain, because the exact conditions required for jet formation are not well-known. We therefore divide our discussion between events that have jets and those that do not.




\subsection{Counterparts of Events with Jet Formation}

Even if a jet is produced, the observational signature can vary greatly depending on whether the jet breaks out of the envelope. For this to occur, a substantial fraction of the envelope must be removed. This is consistent with the correlation of long gamma-ray bursts with Type Ic supernovae \citep{wb06}, which do not have signatures of hydrogen or helium. If there is a sufficiently large hydrogen envelope to prevent the jet from cleanly breaking out, a jet-powered \mbox{Type II} supernovae may result instead. Such an event might look like SN 2010jp \citep{smi12}, which showed evidence for a bipolar outflow. In addition, it had a low output of radioactive $^{56}$Ni ($\lesssim0.003M_\odot$), consistent with appreciable fallback, and signs of low metallicity, consistent with less mass loss \citep{heg03}. A jet-driven explosion was also suggested for SN 2001ig based on spectropolarimetry and line-profile evidence \citep{mau07,sil09}. Other studies have attempted to use polarization to infer the amount of asymmetry in the supernova (which could be driven by a jet), but this is less conclusive \citep{leo00,leo01}. It is enticing though that the amount of asymmetry seems to be inversely related to the hydrogen envelope mass \citep[][and references therein]{cho11}, which is consistent with the generally expected picture. If the hydrogen envelope buries all signs of a jet, even a weak amount of subsequent hydrogen-recombination may produce a luminous red nova-like event \citep{kul07,bon09,tho09}, as has been speculated by \citet{qk12}.

\subsection{Events without Jet Formation}

If the majority of the material just falls into the BH or if there is little $^{56}$Ni production, the electromagnetic signature may be very weak, producing an ``unnova'' \citep{koc08}. The prospect of identifying dying stars that cannot be observed with traditional methods would be an exciting use of GWs. On the other hand, there is currently little evidence for such events, and no apparent need of them to explain discrepancies between rates of different supernova types versus a standard initial mass functions \citep{smi11}. Furthermore, the nearby Type II-L SN 1979C \citep{pan80} clearly had a bright, explosive display, but some have argued that it left a BH remnant \citep{pat11}. A dim or nonexistent event also seems to be disfavored on physical grounds. The outer envelope of a red supergiant is very loosely bound, and just a weak shock will eject it and produce a signature from hydrogen recombination \citep{des10}, even if all the radioactive material falls into the BH. This would argue that a supernova with fallback may produce a Type II-P like light curve, even if it is relatively weak. Again, this might lead to a connection with some subset of the luminous red novae.

\subsection{Rate Estimates}

Since it has been difficult to definitively connect any electromagnetic transients with fallback accretion and BH formation, the rate of such events, especially within the $\approx17\ {\rm Mpc}$ distance needed for GW detection, is uncertain. Long gamma-ray bursts may be ideal sources, since GWs from fallback may be a precursor to jet formation. Such a detection would provide important evidence that some long gamma-ray bursts are powered by accretion versus models that favor young magnetars \citep[][and references therein]{met11}. Unfortunately, the closest recorded long gamma-ray burst was at a distance of $37\ {\rm Mpc}$ \citep{gal98}, making the prospect of a nearby long gamma-ray burst unlikely.
Based on rate estimates from \citep{chapman}, we expect $\unit[3]{kyr^{-1}}$ within $\unit[17]{Mpc}$.

On the other hand, $\sim1-2$ supernovae typically occur within $17\ {\rm Mpc}$ each year; see Table~\ref{tab:sne}. Most of these nearby events are the common Type II-P supernovae, which are perhaps the least likely to have fallback accretion, although there are a few other types. Nevertheless, given the uncertainties in the electromagnetic signature from fallback, events like these would be worth checking with next generation gravitational wave interferometers. The Hubble Space Telescope can detect progenitors of these events in pre-explosion imaging out to $\approx30\ {\rm Mpc}$ \citep{sma09b}. So in principle, we might expect that any supernova that has GWs detectable from fallback will therefore have an identified progenitor. In practice, only $\sim1/3$ of the time is the archival imaging available for the location of the nearby supernovae. Also problematic is that there still has never been a progenitor star identified for a Type Ib or Ic supernovae, which remains an outstanding puzzle.

\begin{deluxetable}{c c c c}
  \tablecolumns{4} \tablewidth{0pt}
  \tablecaption{Recent Nearby Core-collapse Supernovae\tablenotemark{a}}
  \tablehead{
    \colhead{Name} & 
    \colhead{Type} & 
    \colhead{Galaxy} & 
    \colhead{Distance (Mpc)} }
  \startdata
  2012aw & II-P & M95 & 10 \\
  2012A & II-P & NGC 3239 & 8 \\
  2011ja & II-P & NGC 4945 & 3.6 \\
  2011dh & IIb & M51 & 7.1 \\
  2009dd & II & NGC 4088 & 15.8 \\
  2008bk & II & NGC 7793 & 3.9 \\
  2008ax & Ib & NGC 4490 & 15 \\
  2007it & II & NGC 5530 & NA\tablenotemark{b} \\
  2007gr & Ib/c & NGC 1058 & 8.4
   \enddata
   \tablenotetext{a}{Summarized from lists of the Astronomy Section, Rochester Academy of Science (ASRAS), http://www.rochesterastronomy.org/snimages/}
   \tablenotetext{b}{The distance for this event has not been reported. Nevertheless, since it was the brightest supernova of 2007 (apparent magnitude), it was likely within the $\approx17\ {\rm Mpc}$ that we require.}
   \label{tab:sne}
\end{deluxetable}

Although much more speculative, if there is fallback accretion associated with some subset of luminous red novae, this could provide an important increase in the rate of fallback GW detections by $\approx50\%$ or more since these events are relatively nearby \citep{tho09}. Since luminous red novae are often shrouded in dust, they can be difficult to study and may even be missed electromagnetically in some instances. GWs provide a method to potentially detect and learn about these mysterious transients.

The rate of events without any electromagnetic signature is the most difficult to estimate. Monitoring data for M81 from the Large Binocular Telescope formally set the rate as $\lesssim80$ times that of normal SNe at 90\% confidence \citep{pri08}, which is not especially limiting. Perhaps more constraining, the similarity of massive star formation rates and supernova rates and the nondetection of a diffuse supernova neutrino background both indicate that the rate of events without electromagnetic signatures cannot significantly exceed the rate of observed core-collapse supernovae \citep{hb06}. Even though a systematic survey is necessary to provide stringent limits on such events \citep{koc08}, the prospect that GWs could help find a missing class of dying stars is worth exploring in more detail.




\section{Discussion and Conclusion}
\label{sec:end}
We investigated the production of GWs in a subset of supernovae where a NS is produced in the core, but which subsequently collapses to a BH after $\approx\unit[30-3000]{s}$ of fallback accretion. Using semi-analytic models for the spin evolution of the NS, we estimate typical GW frequencies of $\approx\unit[700-2400]{Hz}$ as well as the corresponding strain amplitudes. 
We estimate that such events will be detectable by Advanced LIGO out to $\approx\unit[17]{Mpc}$.
Given the messy process of fallback accretion, we argue that this figure is much more robust than estimates made with an overly optimistic matched-filtering technique. Finally, we discuss a number of electromagnetic events that may be associated with this unique GW signal. Although such connections are still very uncertain, there are likely $\sim1-2$ supernovae events per year that will be worth searching for fallback GWs.

In this initial work a number of simplifying assumptions were made in the GW model. An important focus of future studies is to investigate more realistic progenitors. 
Future improvements may include: using realistic fallback rates from numerical models, resolving the time-dependent accretion disk that mediates the flow of matter onto the NS, and better understanding what value of $\beta_c$ is appropriate for a NS being spun up by fallback. Furthermore, a better survey of the range of fallback rates and expected timescales should be conducted, especially in the case of stellar models that may have been ignored because they do not produce long gamma-ray bursts. Binary interactions may also play an important role in generating the needed rotation rates \citep[for example, see][]{wh12}, a fact that is maybe not surprising given that the majority of massive stars are in close binary systems \citep{chi12}.

In the event of a detection, there are a number of interesting things we might learn from the GW signature. For a given frequency, there is a one-to-one correspondence with the density of the NS, given constraints on the polytropic index $n$ and $\beta_c$ via \mbox{equation (\ref{eq:omega}).} Therefore determining the frequency at the moment before collapse to a BH would directly constrain the density at the maximum mass of a NS, an important constraint on its equation of state \citep{lp07}. The main uncertainty is in $\beta_c$. We have argued that $\beta_c\approx\beta_{\rm sec}$ is the most likely possibility, but even if $\beta_c\approx0.1-0.15$, this would only introduce a $\approx30\%$ error in the density assuming the frequency is precisely measured. Nevertheless, this highlights the importance of determining $\beta_c$ in future, theoretical work.
\\


We thank Christian Ott for detailed comments. We also thank Lee Lindblom for helpful discussions on nonaxisymmetric instabilities of rotating NSs, and Andrew MacFadyen for feedback on the observational impact of fallback accretion during supernovae. A.L.P. was supported through NSF grants AST-0855535 and PHY-1069991, and by the Sherman Fairchild Foundation. E.T. was supported by NSF grant PHY-0758035.
This is LIGO document \#P1200084.


\end{document}